 \documentclass[preprint,5p,times,twocolumn]{elsarticle}

\usepackage{amssymb}
\usepackage{amsmath}
\usepackage[margin=0pt,font=small,labelfont=bf,labelsep=colon]{caption}

\usepackage{algorithmic}
\usepackage{algorithm}

\usepackage{listings}
\usepackage{color}
\usepackage{textcomp}
\definecolor{listinggray}{gray}{0.9}
\definecolor{lbcolor}{rgb}{0.9,0.9,0.9}
\journal{Computer Physics Communications}

\begin{document}

\begin{frontmatter}


\author{T. Siro\corref{cor1}}
\ead{topi.siro@aalto.fi}
\cortext[cor1]{Corresponding author}
\author{A. Harju}


\title{Exact diagonalization of the Hubbard model on graphics processing units}


\address{Aalto University School of Science, P.O. Box 14100, 00076 Aalto, Finland}

\begin{abstract}
We solve the Hubbard model with the exact diagonalization method on a graphics processing unit (GPU). We benchmark our GPU program against a sequential CPU code by using the Lanczos algorithm to solve the ground state energy in two cases: a one-dimensional ring and a two-dimensional square lattice. In the one-dimensional case, we obtain speedups of over 100 and 60 in single and double precision arithmetic, respectively. In the two-dimensional case, the corresponding speedups are over 110 and 70.  

\end{abstract}


\end{frontmatter}


\section{Introduction}
\label{intro}
The Hubbard model was introduced in the 1960s to describe interacting electrons in a solid\cite{Kanamori,Gutzwiller,Hubbard}. It has since been the subject of extensive study and is still a source of interesting physics\cite{spinliquid}. It is perhaps the simplest model to display many of the essential features of electron correlations, such as ferromagnetism and conductor-Mott-insulator transition.

In the Hubbard model, the solid is described by a fixed lattice, where electrons can hop from one lattice site to another. The electrons are always bound to an atom, such that their wave functions are vectors whose components squared are the probabilities of finding the electron at the corresponding lattice site. Interactions take place only between electrons that are residing on the same site. 

The Hamiltonian can be written as
\begin{eqnarray}
\displaystyle H &=& H_{hop}+H_{int} \\
&=& -t\sum_{<ij>}\sum_{\sigma=\uparrow,\downarrow}(c_{i,\sigma}^{\dagger}c_{j,\sigma}+h.c)+U\sum_{i}n_{i,\uparrow}n_{i,\downarrow},
\label{eq:H}
\end{eqnarray}
where $<ij>$ denotes a sum over neighboring lattice sites, $c_{i,\sigma}^{\dagger}$ and $c_{i,\sigma}$ are the creation and annihilation operators which respectively create and annihilate an electron at site $i$ with spin $\sigma$, and $n_{i,\sigma}=c_{i,\sigma}^{\dagger}c_{i,\sigma}$ counts the number of such electrons. $U$ is the interaction energy and $t$ is the hopping amplitude. The creation and annihilation operators obey the typical anticommutation rules for fermions,
\begin{equation}
\left\{ c_{i\sigma}^{\dagger},c_{j\tau}\right\} =\delta_{ij}\delta_{\sigma\tau}
\quad{} \textnormal{and} \quad{}
\left\{ c_{i\sigma}^{\dagger},c_{j\tau}^{\dagger}\right\} =\left\{ c_{i\sigma},c_{j\tau}\right\} =0,
\end{equation} which means that there are four possible occupations for a lattice site: either it is empty, has one up electron, one down electron or one of each.

An important property of the Hamiltonian is that the numbers of both up and down electrons are separately conserved. This is convenient because it allows one to fix the number of up and down electrons and thus restrict to a subspace of the whole Hilbert space. 

Despite the model's simplicity, an analytic solution is only available in one dimension, and it was found by Lieb and Wu in 1968\cite{LiebWu}. In general, computational methods are required. While both terms in the Hamiltonian are easy to diagonalize separately, their sum is highly nontrivial. One method to numerically solve the Hubbard model is exact diagonalization. The idea is to simply calculate the matrix elements in a suitable basis and then diagonalize the resulting matrix. The obvious downside of this approach is that the number of lattice sites and particles that can be considered is quite low due to the very rapid growth of the dimension of the Hamiltonian matrix as a function of the system size. However, the major advantage is that the results are exact up to numerical accuracy, which makes exact diagonalization well-suited to situations where a perturbative solution is not possible. It can also be used to test the reliability of other, approximative methods by comparing their results with the exact ones.

\section{Exact diagonalization}
\subsection{The Hamiltonian}
\label{Hamiltonian}

To calculate the matrix elements of the Hubbard Hamiltonian in Equation (\ref{eq:H}), we choose a simple basis where the lattice sites are numbered from 0 upward and the basis states correspond to all the ways of distributing the electrons in the lattice. For example, if we have $N_{s}=4$ lattice sites, $N_{\uparrow}=2$ spin up electrons and $N_{\downarrow}=3$ spin down electrons, then
\begin{equation}
c_{0\uparrow}^{\dagger}c_{2\uparrow}^{\dagger}c_{0\downarrow}^{\dagger}c_{2\downarrow}^{\dagger}c_{3\downarrow}^{\dagger}\left|\mathcal{O}\right\rangle \label{eq:basisstate}\end{equation}
is one basis state. In state (\ref{eq:basisstate}), the up electrons reside on sites 0 and 2 and the down electrons on sites 0, 2 and 3. The empty lattice into which the electrons are created is denoted by $\left|\mathcal{O}\right\rangle$. To resolve any ambiguities arising from different orderings of the creation operators in Equation (\ref{eq:basisstate}), we define that in the basis states all spin up operators are to the left of all spin down operators, and site indices are in ascending order.

The dimension of the Hamiltonian matrix is equal to the number of ways of distributing $N_{\uparrow}$ spin up electrons and $N_{\downarrow}$ spin down electrons into $N_{s}$ lattice sites, i.e.
\begin{equation}
\dim H = \binom{N_{s}}{N_{\uparrow}}\binom{N_{s}}{N_{\downarrow}}.
\end{equation}
The size of the basis grows extremely fast. For example, in the half-filled case where $N_{\uparrow} = N_{\downarrow} = N_{s}/2$, for 12 sites $\dim H=853776$, for 14 sites $\dim H\approx11.8\times10^{6}$ and for 16 sites $\dim H\approx166\times10^{6}$. In addition, the matrices are very sparse, because the number of available hops, and thus the number of nonzero elements in a row, grows only linearly while the size of the matrix grows exponentially.

To form the Hamiltonian, we need to label and order the basis states. A convenient way to do this is to represent each state with binary numbers such that occupied and unoccupied sites are denoted by 1 and 0, respectively. For example, the state in (\ref{eq:basisstate}) becomes
\begin{equation}
c_{0\uparrow}^{\dagger}c_{2\uparrow}^{\dagger}c_{0\downarrow}^{\dagger}c_{2\downarrow}^{\dagger}c_{3\downarrow}^{\dagger}\left|\mathcal{O}\right\rangle \rightarrow\underbrace{(0101)}_{up}\times\underbrace{(1101)}_{down}.\label{eq:binaryrep}\end{equation}
Note that in our convention site indices run from right to left in the binary number representation.

A simple way to order the basis states is to do it according to the size of the binary number that represents the state.Using this scheme, if we index the states by $J$, the conversion
from the binary representation is given by 
\begin{equation}
J=i_{\uparrow}\dbinom{N_{s}}{N_{\downarrow}}+i_{\downarrow},\label{eq:J-formula}
\end{equation}
where $i_{\uparrow}$ and $i_{\downarrow}$ are the positions of the
up and down configurations in an ordered list, starting from 0, of
all $N_{s}$-bit numbers with $N_{\uparrow}$ and $N_{\downarrow}$
bits set, respectively. To clarify, for example in (\ref{eq:binaryrep}),
the possible up configurations, in order, are 0011, 0101, 0110, 1001,
1010 and 1100, so 0101 is the second configuration, and $i_{\uparrow}=1$.
Similarly, 1101 is the third 4-bit number with 3 bits set, so $i_{\downarrow}=2$.
Thus, we get
\begin{equation}
J=1\times\binom{4}{3}+2=6,
\end{equation}
which is confirmed by Table \ref{tab:A-scheme-for}.%
\begin{table}[t]
\begin{centering}
\begin{tabular}{|c|c|c|c|c|}
\hline 
$\uparrow$ & $\downarrow$ & $i_{\uparrow}$ & $i_{\downarrow}$ & $J$\tabularnewline
\hline
\hline 
0011 & 0111 & 0 & 0 & 0\tabularnewline
\hline 
0011 & 1011 & 0 & 1 & 1\tabularnewline
\hline 
0011 & 1101 & 0 & 2 & 2\tabularnewline
\hline 
0011 & 1110 & 0 & 3 & 3\tabularnewline
\hline 
0101 & 0111 & 1 & 0 & 4\tabularnewline
\hline 
0101 & 1011 & 1 & 1 & 5\tabularnewline
\hline 
0101 & 1101 & 1 & 2 & 6\tabularnewline
\hline 
0101 & 1110 & 1 & 3 & 7\tabularnewline
\hline 
0110 & 0111 & 2 & 0 & 8\tabularnewline
\hline 
$\vdots$ & $\vdots$ & $\vdots$ & $\vdots$ & $\vdots$\tabularnewline
\hline 
1100 & 1101 & 5 & 2 & 22\tabularnewline
\hline 
1100 & 1110 & 5 & 3 & 23\tabularnewline
\hline 
\end{tabular}
\par\end{centering}

\caption{A scheme for labelling the basis states for $N_{s}=4$, $N_{\uparrow}=2$,
$N_{\downarrow}=3$. States are ordered first according to the up
spin configuration (first column) and then according to the down spin
configuration (second column), in ascending order. \label{tab:A-scheme-for}}
\end{table}

Forming the Hamiltonian matrix is now straightforward. The interaction part $H_{int}$ is diagonal, and it essentially just counts the number of doubly occupied lattice sites and increases the energy by $U$ for each instance. The matrix elements of the hopping part, $H_{hop}$, are $\pm t$ between basis states that can be reached from each other by a hop of a single electron, and vanish otherwise. 

For example, if we have a one-dimensional lattice with periodic boundaries, from Table \ref{tab:A-scheme-for}, we see that
\begin{equation}
|\left\langle 2\right|H_{hop}\left|6\right\rangle| = t \quad{} \textnormal{and} \quad{} \left\langle 3\right|H_{hop}\left|6\right\rangle = 0
\end{equation}
because $\left|6\right\rangle$ can be reached from $\left|2\right\rangle$ when the up electron at site 1 hops to site 2. From $\left|3\right\rangle$ to $\left|6\right\rangle$ it takes two hops so the matrix element vanishes. Because of the binary number representation of the basis states, in the computer these calculations can be conveniently done with integers and bitshift operations.

The signs of the nonzero matrix elements are determined by the anticommutation relations of the creation and annihilation operators. An extra minus sign is picked up for each electron of the same spin that is hopped over. So in the one-dimensional case, if the hop is over the periodic boundary and the total number of electrons of the same spin is even, the matrix element changes sign. For example,  
\begin{equation}
\left\langle 6\right|H_{hop}\left|22\right\rangle = t \quad{} \textnormal{and} \quad{} \left\langle 0\right|H_{hop}\left|3\right\rangle = -t
\end{equation}
because there is an even number of up spins and an odd number of down spins (note the minus in the Hamiltonian). Note that the method is completely general and applies to any kind of lattice and any number of electrons. In a general lattice, the sign is determined by the number of electrons of the same spin residing in lattice sites whose labels are between the labels of the origin and the destination of the hop.

\subsection{The Lanczos algorithm}
\label{lanczos}

It is apparent that fully diagonalizing the Hamiltonian using standard methods is only practical for very small systems. Fortunately, we are usually mostly interested in the ground state and the lowest excited states. A well known method for accurately approximating the lowest eigenvalues and eigenstates of sparse matrices is the Lanczos algorithm\cite{itermethods}. 

The idea of the Lanczos algorithm is to project the Hamiltonian onto an orthogonalized basis in a Krylov subspace, defined by
\begin{equation}
\mathcal{K}_{m}(f,H)=\text{span}(f,Hf,H^{2}f,\dots,H^{m-1}f),
\end{equation}
where $f$ is a random starting vector and $m$ is the dimension of the subspace. The result of the Lanczos procedure is a tridiagonal matrix, i.e. one with nonzero elements only on the main diagonal and the first sub- and superdiagonals. The dimension of the Krylov space does not have to be known beforehand. Instead, one can check for convergence after each iteration, for example by calculating the difference in the ground state energy from the previous iteration. The lowest (and highest) eigenvalues accurately approximate the corresponding eigenvalues of $H$ already for $m \ll \dim H$ .

\begin{algorithm}[t]                      
\caption{The Lanczos algorithm \cite{itermethods}.}          
\label{alg1}                           
\begin{algorithmic}[1]  
\REQUIRE a random initial vector $f_1$ of norm 1
 \STATE $b_1 \leftarrow 0$
 \STATE $f_0 \leftarrow 0$
\FOR{$j=1$ to $m$}
     \STATE $q_j \leftarrow Hf_{j}-b_{j}f_{j-1}$
     \STATE $a_j \leftarrow q_{j}^{\dagger}f_j$
     \STATE $q_j \leftarrow q_{j}-a_{j}f_j$
     \STATE $b_{j+1} \leftarrow \sqrt{q_{j}^{\dagger}q_j}$. 
     \IF{$b_{j+1}=0$} 
     \STATE Stop
     \ENDIF
     \STATE $f_{j+1} \leftarrow q_{j}/b_{j+1}$
\ENDFOR
\end{algorithmic}
\end{algorithm}


One way to write the Lanczos algorithm is, in pseudocode, given by Listing 1. The algorithm generates the so called Lanczos basis, $\{f_{1},f_{2},\dots,f_{m}\}$,
in the Krylov space, and the projection of $H$ in this basis is given
by the generated constants $a_{j}$ and $b_{j}$ as\[
\label{Tmatrix}
T=\left(\begin{array}{ccccc}
a_{1} & b_{2} & 0 & \cdots & 0\\
b_{2} & a_{2} & b_{3} & \ddots & \vdots\\
0 & \ddots & \ddots & \ddots & 0\\
\vdots & \ddots & b_{m-1} & a_{m-1} & b_{m}\\
0 & \cdots & 0 & b_{m} & a_{m}\end{array}\right).\]
Since $T$ is tridiagonal, symmetric and typically much smaller than
$H$, its eigenvalues are easy to calculate with standard methods.

Since memory is a scarce resource, especially on the GPU, we would like to use as few vectors as possible. Superficially, it seems like three vectors are needed for the Lanczos algorithm because of the three-term recurrence relation on line 4 of Algorithm \ref{alg1}:
\begin{equation}
q_j \leftarrow Hf_{j}-b_{j}f_{j-1}.
\end{equation}
However, we can manage with only two vectors by splitting this into
\begin{align}
f_{j-1} &\leftarrow -b_{j}f_{j-1}\label{split1}\\
f_{j-1} &\leftarrow f_{j-1} + Hf_{j}\tag{\ref{split1}'}
\end{align}
and then renaming $f_{j-1}$ to $q_j$. 

\section{GPU computing}
\subsection{Introduction}

Graphics processing units (GPU), originally developed to aid the central processing unit (CPU) in rendering graphics, have evolved into powerful computational engines, capable of general purpose calculations. In recent years, they have been increasingly used in a variety of scientific disciplines, including physics, to speed up computations that benefit from the architecture of the GPU.

The GPU is quite different than the CPU. Simply put, while the CPU performs few concurrent tasks quickly, the GPU executes a very large number of slower tasks simultaneously, i.e. in parallel. Although modern CPUs typically consist of multiple cores, allowing parallel computation to some extent, the scale of parallelization in the GPU is orders of magnitude larger, up to tens of thousands of simultaneous threads in modern cards. 

To benefit from GPUs, the problem has to be suitable for large scale parallelization. In addition, specifics of the GPU architecture need to be taken into account. For example, to get a performance gain, the program should have high arithmetic intensity, defined as the number of arithmetic operations divided by the number of memory operations. This is because accesses to the memory have a high latency and it is therefore desirable to hide this latency with calculations. Also, data transfer between the CPU and the GPU through the PCI Express bus is slow, and should therefore be minimized.

\subsection{CUDA}

Compute Unified Device Architecture (CUDA) is a parallel computing architecture by the GPU manufacturer NVIDIA. It allows programmers to utilize GPUs with CUDA capability in general purpose computation through the use of an extension of the C programming language.

CUDA programs consist of the main program that runs on the CPU (\emph{the host}), and special functions called kernels that run on the GPU (\emph{the device}). Since the GPU has a SIMD architecture, kernels are written from the viewpoint of a single thread. Threads are organized into groups that are called blocks.

When a kernel is launched from the host code, the programmer specifies the number of blocks and the number of threads per block in the kernel call. Each thread has its identification number within the block stored in an internal variable threadIdx, which can be one-, two- or three-dimensional. Similarly, the block ID is stored in a variable called blockIdx. With these ID numbers the programmer can assign the threads to different data.

Threads can be synchronized within a block, but blocks have to be able to execute independently. Within a block, threads have a common fast memory space called shared memory that can be used as a user-managed cache. Optimally, one would like to read the relevant data from the slow global memory into the fast shared memory, perform the calculation there and write the result back to the global memory. In the latest generation of NVIDIA GPUs, called Fermi, there are also automatic L1 and L2 caches for global memory accesses. For a more thorough overview of CUDA, we refer to Ref. \cite{CUDAguide}

\section{GPU implementation}
\subsection{Overview}

In the following subsections, we will describe the details of our GPU implementation. The general structure of the program is that of a typical GPU application: first, we prepare the data, such as the Hamiltonian matrices, on the host. Then, we allocate memory on the device and transfer the data to the global memory of the GPU. Computations are then done on the GPU by launching the necessary kernels, and finally the results are transferred back to the host.

Our application runs on a single GPU. A multi-GPU implementation, while certainly feasible, would not offer significant benefits from a physical point of view. This is because the required amount of memory grows so rapidly with the system size that even with a large number of GPUs, we would not have access to significantly larger systems compared to a single GPU. Additionally, since the communication between GPUs currently has to take place through the host system, it might be difficult to achieve good multi-GPU performance because of the large amount of communication needed in the Hamiltonian times vector operation. 

\subsection{Storing the Hamiltonian}

The overwhelmingly most time-consuming part of the Lanczos algorithm for large systems is the operation on a vector with the Hamiltonian on line 4 of Algorithm \ref{alg1}. For small matrices, this could be implemented as simple matrix-vector multiplication. Indeed, there have been previous GPU implementations of the Lanczos algorithm relating to graph bisection and image segmentation\cite{LanczosBisect} as well as latent semantic analysis\cite{Lanczos1}. In these works, the matrix sizes have been small enough to allow a direct computation of the matrix-vector product.

However, as discussed in Subsection \ref{Hamiltonian}, our Hamiltonian matrix becomes very large already for systems with over 10 lattice sites. The newest Fermi generation of NVIDIA graphics cards have up to 6 GB of memory, so there is no hope of storing the full Hamiltonian in the memory of the GPU. With the interaction part this is not a problem because it can be easily generated on the fly, but with the hopping part we would prefer to have it precomputed.

To compress $H_{hop}$, we first split it into a Kronecker sum of up and down hopping Hamiltonians, which operate only on electrons with the corresponding spin:
\begin{equation}
H_{hop}=H_{\uparrow}\oplus H_{\downarrow}=H_{\uparrow}\otimes I_{\downarrow}+I_{\uparrow}\otimes H_{\downarrow},\label{eq:H-Kroneckersum}
\end{equation}
where $I_{\sigma}$ is the identity operator for electrons with spin  $\sigma$
and $\otimes$ denotes the Kronecker product of matrices, which corresponds
to the tensor product of linear maps, and is defined as follows: If
$A$ is a $m$-by-$n$ matrix with elements $a_{ij},$ then the Kronecker
product of $A$ and another (arbitrary sized) matrix $B$ is a block
matrix defined by \[
A\otimes B=\left(\begin{matrix}a_{11}B & \dots & a_{1n}B\\
\vdots & \ddots & \vdots\\
a_{m1}B & \dots & a_{mn}B\end{matrix}\right).\]

Further, because $H_{\uparrow}$ and $H_{\downarrow}$ are still very sparse, we can store them efficiently by using a standard sparse matrix format. We choose the ELL format\cite{spmv}, in which the sparse matrix is compressed horizontally into a dense matrix. The width of the resulting matrix is determined by the maximum number of nonzero elements per row in the original matrix. Another matrix of the same size contains the column indices of the elements. ELL is well-suited for matrices whose number of nonzero elements per row does not change very much because that means that little space is wasted to padding. This is the case with  $H_{\uparrow}$ and $H_{\downarrow}$.

\begin{figure}[t]
\[
A=\left(\begin{array}{cccc}
5 & 1 & 0 & 0\\
0 & 2 & 7 & 3\\
4 & 0 & 6 & 0\\
0 & 9 & 8 & 0\end{array}\right)\]
\[
\Downarrow\]
\[
\text{data}=\left(\begin{array}{ccc}
5 & 1 & *\\
2 & 7 & 3\\
4 & 6 & *\\
9 & 8 & *\end{array}\right)\quad\text{indices}=\left(\begin{array}{ccc}
0 & 1 & *\\
1 & 2 & 3\\
0 & 2 & *\\
1 & 2 & *\end{array}\right)\]
\[
\Downarrow\]
\begin{eqnarray*}
\text{data} & = & (5,2,4,9,1,7,6,8,*,3,*,*)\\
\text{indices} & = & (0,1,0,1,1,2,2,2,*,3,*,*)\end{eqnarray*}

\caption{The ELL format produces two smaller matrices from the initial sparse
matrix. In practice, these will be converted to vectors in column-major
order. The stars denote padding and they are set to zero.}
\label{ELL}
\end{figure}

As shown in Fig. \ref{ELL}, the ELL formatted matrices are stored in the memory column-wise. This is because in the matrix-vector product each dot product of a row and the vector is done in parallel. Thus, in the first iteration, the threads access the elements of the first column, in the second iteration the second column etc. When the matrix is stored column-wise, these memory accesses are contiguous, resulting in better memory bandwidth.

After the splitting of $H_{hop}$ into $H_{\uparrow}$ and $H_{\downarrow}$ and storing them with the ELL format, the size of the Hamiltonian is no longer a problem. Instead, the memory consumption is almost solely determined by the number of stored state vectors. The complication is that operating on a vector with $H_{hop}$ is no longer simple matrix-vector multiplication.

\subsection{The kernel}

The core of our GPU implementation of the exact diagonalization procedure is the kernel that computes the product of the Hamiltonian and a state vector. It naturally consists of three different parts: operating with the up part of $H_{hop}$, the down part of $H_{hop}$ and $H_{int}$. To avoid unnecessary memory accesses, it is beneficial to do them all in the same kernel. 

To see what the spin up hopping Hamiltonian does to a state, it is useful to think of the vector as consisting of $\dim H_{\uparrow}$ subvectors of length $\dim H_{\downarrow}$. The basis states within each subvector have the same up spin configuration. It is then straightforward to show that operating with the spin up hopping Hamiltonian is just like normal matrix-vector multiplication on the vector of subvectors:
\begin{equation}
 (H_{\uparrow}\otimes I_{\downarrow})x=H_{\uparrow}\left(\begin{array}{c}
\mathbf{x}^{(0)}\\
\mathbf{x}^{(1)}\\
\vdots\\
\mathbf{x}^{(\dim H_{\uparrow}-1)}\end{array}\right).
\end{equation}

\begin{algorithm}[t]                     
\caption{The kernel pseudocode for operating with the spin up hopping Hamiltonian}          
\label{Hup}                           
\begin{algorithmic}[1] 
\REQUIRE blockID \COMMENT{the block index}
\REQUIRE sv \COMMENT{the subvector index}
\REQUIRE id \COMMENT{the thread index within the subvector}
\REQUIRE blockID $<$ dimUp * blocksPerSubvector
\REQUIRE id $<$ dimDn  

\STATE sum $\leftarrow 0$
\STATE
\IF{threadIdx.x $<$ numcolsUp}         
          \STATE Ax$_s$[threadIdx.x]$ \leftarrow$ AxUp[threadIdx.x*dimUp + sv]
          \STATE Aj$_s$[threadIdx.x]$ \leftarrow$ AjUp[threadIdx.x*dimUp + sv]
\ENDIF     

\STATE syncthreads
	  
\STATE
\FOR{$i=0$ to numcolsUp}         
      \IF{Ax$_s \neq 0$}   
          \STATE sum $\leftarrow$ sum + Ax$_s$[i] * x[Aj$_s$[i] * dimDn + id]
      \ENDIF 
\ENDFOR
\end{algorithmic}
\end{algorithm}

\begin{figure}[t]
\[
x=\left(\begin{array}{c}
x_{0}\\
x_{1}\\
\vdots\\
x_{\dim H_{\downarrow}-1}\\
\hline x_{\dim H_{\downarrow}}\\
x_{\dim H_{\downarrow}+1}\\
\vdots\\
x_{2\dim H_{\downarrow}-1}\\
\hline \vdots\\
x_{\dim H_{\uparrow}\dim H_{\downarrow}-1}\end{array}\right)\equiv\left(\begin{array}{c}
\mathbf{x}^{(0)}\\
\mathbf{x}^{(1)}\\
\vdots\\
\mathbf{x}^{(\dim H_{\uparrow}-1)}\end{array}\right)\]

\caption{The state vectors can be thought to consist of subvectors of constant up
spin configuration.\label{fig:slots} }

\end{figure}

Implementing this on the GPU is rather straightforward because we are essentially just copying parts of the input vector to other parts in the output vector. Therefore, we will launch $\dim H_{\downarrow}$ threads per row that calculate the dot product of that row with $x$. The kernel code can be seen in Algorithm \ref{Hup}. 

The kernel receives the two ELL formatted matrices as inputs, and inside the kernel,  Up and Dn in variable names refer to $H_{\uparrow}$ and $H_{\downarrow}$, respectively. The variables dim contain the dimensions of the Hamiltonians, and numcols have the number of columns of the ELL representations. The ELL formatted matrices are in the arrays Ax (the data vector in Figure \ref{ELL}) and Aj (the indices vector in Figure \ref{ELL}).

Due to hardware limitations, we might need one or more blocks per subvector in order to have one thread per element of $x$. In the beginning of the kernel, we calculate a couple of helpful variables, the subvector index of the current thread,~sv = blockID / blocksPerSubvector (integer division) and the thread index within the subvector, id = (blockID $\mod$ blocksPerSubvector) $\times$ threadsPerBlock + threadIdx.x.

After discarding any extraneous blocks or threads, we load the elements of the appropriate row of the spin up hopping Hamiltonian into shared memory arrays Ax$_s$ and Aj$_s$ on lines 4 and 5 and synchronize afterwards. Finally, we loop over the Hamiltonian and accumulate the result in the register variable called sum.

Operating with the spin down part of the hopping Hamiltonian is straightforward. We just compute an ordinary matrix-vector product with each subvector: 
\begin{equation}
(I_{\uparrow}\otimes H_{\downarrow})x=\left(\begin{array}{c}
H_{\downarrow}\mathbf{x}^{(0)}\\
H_{\downarrow}\mathbf{x}^{(1)}\\
\vdots\\
H_{\downarrow}\mathbf{x}^{(\dim H_{\uparrow}-1)}\end{array}\right).
\end{equation}
The kernel code for this can be seen in Algorithm \ref{Hdn}.

\begin{algorithm}[t]                      
\caption{The kernel pseudocode for operating with the spin down hopping Hamiltonian}          
\label{Hdn}                           
\begin{algorithmic}[1] 
\REQUIRE blockID \COMMENT{the block index}
\REQUIRE sv \COMMENT{the subvector index}
\REQUIRE id \COMMENT{the thread index within the subvector}
\REQUIRE blockID $<$ dimUp * blocksPerSubvector
\REQUIRE id $<$ dimDn  

\FOR{$i=0$ to numcolsDn}   
      \STATE Aij $\leftarrow$ AxDn[i * dimDn + id];
      \STATE col $\leftarrow$ AjDn[i * dimDn + id];    

         \STATE sum $\leftarrow$ sum + Aij * x[sv * dimDn + col]		
\ENDFOR
\end{algorithmic}
\end{algorithm}

\begin{algorithm}[t]                    
\caption{The kernel pseudocode for operating with the interaction Hamiltonian}          
\label{Hint}                           
\begin{algorithmic}[1] 
\REQUIRE gID \COMMENT{the global thread index within the vector x}
\REQUIRE U \COMMENT{the interaction strength}
\REQUIRE statesUp \COMMENT{the spin up basis states}
\REQUIRE statesDn \COMMENT{the spin down basis states}

\IF{U $\neq 0$}
\STATE doubles $\leftarrow$ statesUp[gID/dimDn] $\&$ statesDn[gID~$\mod$~dimDn]
\STATE intE $\leftarrow$ U * BitCount(doubles)
\STATE sum $\leftarrow$ sum + intE * x[sv * dimDn + id]
\ENDIF
\end{algorithmic}
\end{algorithm}

Finally, we operate with the interaction Hamiltonian (Algorithm \ref{Hint}), which is generated on the fly from the vectors statesUp and statesDn that contain the integers corresponding to the binary representations of the up and down basis states, respectively. First, we generate a binary number where set bits correspond to doubly occupied sites by doing a bitwise and operation between the up and down states on line 2. Then, we count the number of bits set and multiply by $U$ and increment the variable sum by the product of the interaction matrix element and the corresponding element of $x$.

After that, we are done. The final result of the Hamiltonian times $x$ operation has now been accumulated by each thread into sum, and it can be written into an output vector in the global memory of the GPU.

In the above, we have for simplicity assumed that each thread calculates one element of the result vector. A simple, yet perhaps counter-intuitive optimization is to increase the workload of the threads. This is especially helpful in single precision. According to our experiments, best results can be achieved by calculating four elements per thread, i.e. four subvectors per block. While this increases register and shared memory usage, which leads to smaller occupancy, an impressive reduction of up to 50\% in the execution time of the kernel can be obtained compared to calculating only one element per thread.

\section{Performance}

To measure the performance of our GPU implementation, we compared the execution time of the Lanczos algorithm on a Tesla M2070 GPU against a serial CPU program running on a Intel Xeon X5650. The algorithms in the two implementations are identical, including the modification in Equations (\ref{split1}). The validity of our GPU code has been checked by starting the Lanczos algorithm with the same starting vector in both codes, and verifying that the produced matrix elements $a_{j}$ and $b_{j}$ of the tridiagonal matrix in Equation (\ref{Tmatrix}) are identical up to floating point accuracy. In the CPU code, the hopping Hamiltonian is stored using the CSR sparse matrix format\cite{spmv} instead of ELL, for better performance. 

 We measured the time of a single iteration of the loop in Algorithm \ref{alg1} in two cases: a one-dimensional ring and a two-dimensional square lattice with periodic boundary conditions. The 1D system was studied as a function of the system size while keeping the lattice half-filled, i.e. $N_{\uparrow} = N_{\downarrow} = N_{s}/2$. In the 2D system, the lattice size was fixed to $4\times4$ and the particle number was varied, such that there were equal numbers of up and down spin electrons. We need memory for two state vectors in the Lanczos algorithm, which limits us to systems of up to 16 lattice sites. We benchmarked the GPU program with both single and double precision arithmetic. The CPU was running in double precision, but the results for single precision are essentially identical.

\begin{figure}[t] 
   \includegraphics[width=0.48\textwidth]{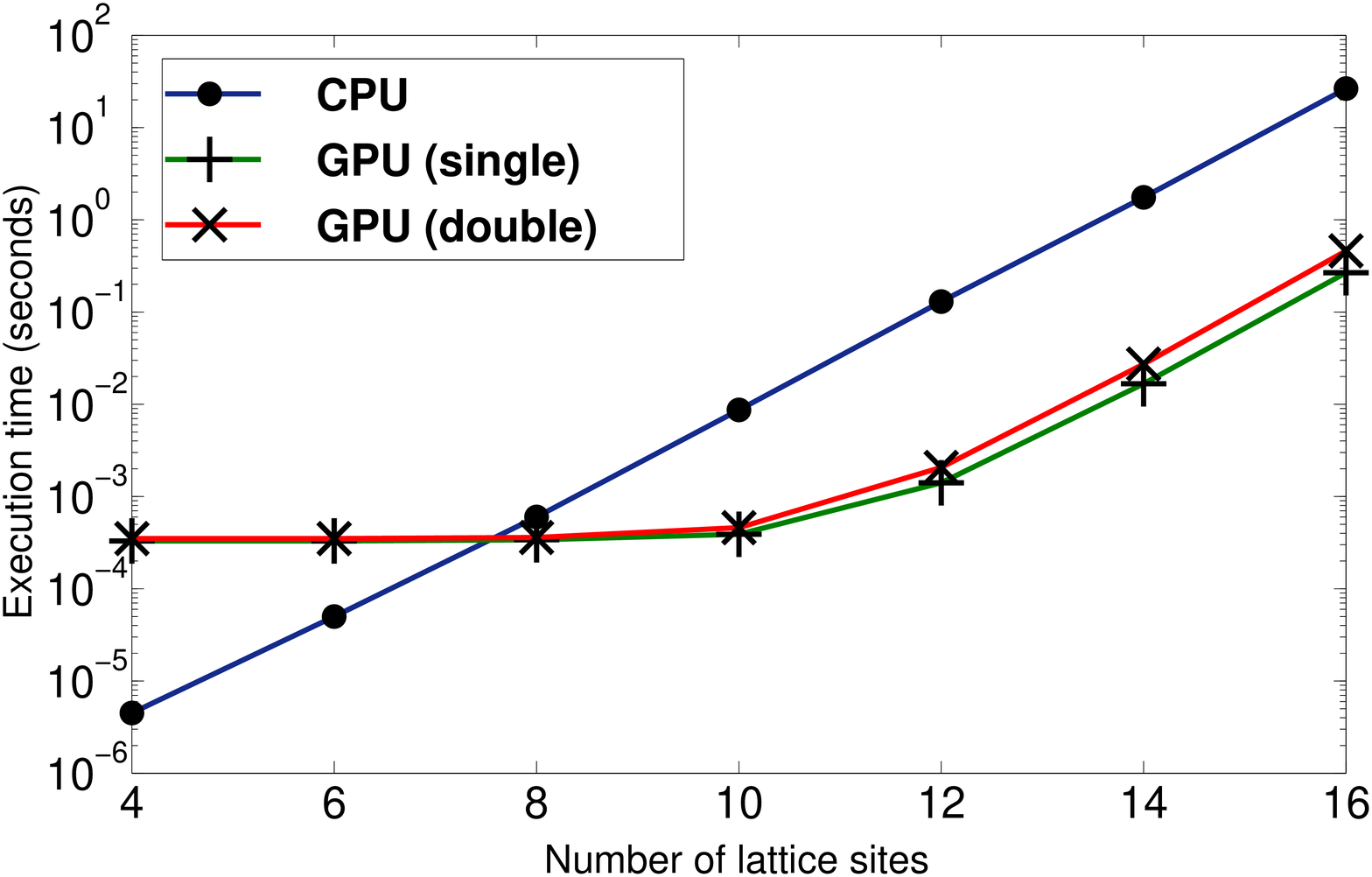} 
   \caption{The execution times of a single iteration of the Lanczos algorithm for the 1D system as a function of the system size.}
   \label{fig:cpuvsgpu}
\end{figure}

The execution times for the 1D system can be seen in Figure \ref{fig:cpuvsgpu}. The CPU shows an expected exponential growth of the execution time as a function of the system size. The GPU, on the other hand, runs in more or less constant time until the system size reaches 10 lattice sites, and for larger systems shows the same exponential scaling as the CPU. 

To understand the behavior, we have profiled our application with the CUDA Visual Profiler. For system sizes below 10, we find that the achieved occupancy of the multiprocessors is very low: 16\% for 8 sites and less for the smaller systems. Also, the profiler reports very low values of below 0.5 for the instructions per clock cycle (IPC) statistic. These suggest that the explanation for the results of  Figure \ref{fig:cpuvsgpu} is that the smaller systems cannot saturate the GPU, and its resources are not fully utilized. Once the GPU is fully occupied, the degree of parallelism reaches its maximum, and any additional workload has to be serialized. This is reflected in IPC values of over 1.5 for systems larger that 10 sites, which is reasonably close to the maximum IPC value of 2. Also, the occupancy is over 60\%, which is enough to hide the memory latency with computation.

\begin{figure}[t] 
 \includegraphics[width=0.48\textwidth]{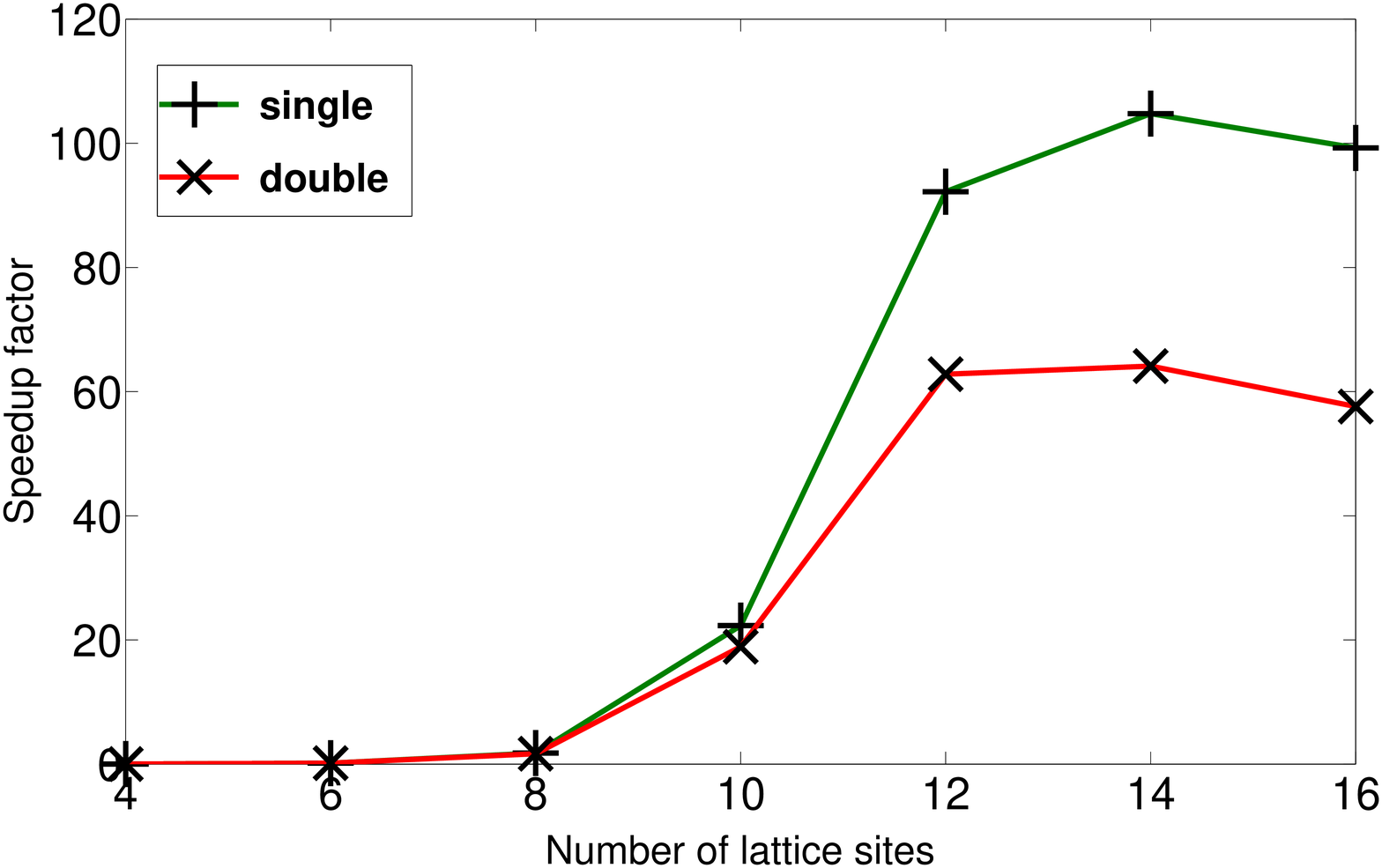} 
   \caption{The speedup factor of the GPU implementation running in single and double precision as a function of the system size for the 1D system. }
   \label{fig:speedup}
\end{figure}

The speedup factor, calculated as the ratio of the CPU and GPU execution times, can be seen in Figure \ref{fig:speedup}. The GPU overtakes the CPU in speed when the system size is 8 lattice sites. After a rapid growth, the speedup stabilizes to around 100 for single precision and 60 for double precision after 12 sites. In both cases, the maximum speedup is achieved with a system size of 14, where the speedups are 105 and 64 for single and double precision, respectively. 

\begin{figure}[t] 
   \includegraphics[width=0.48\textwidth]{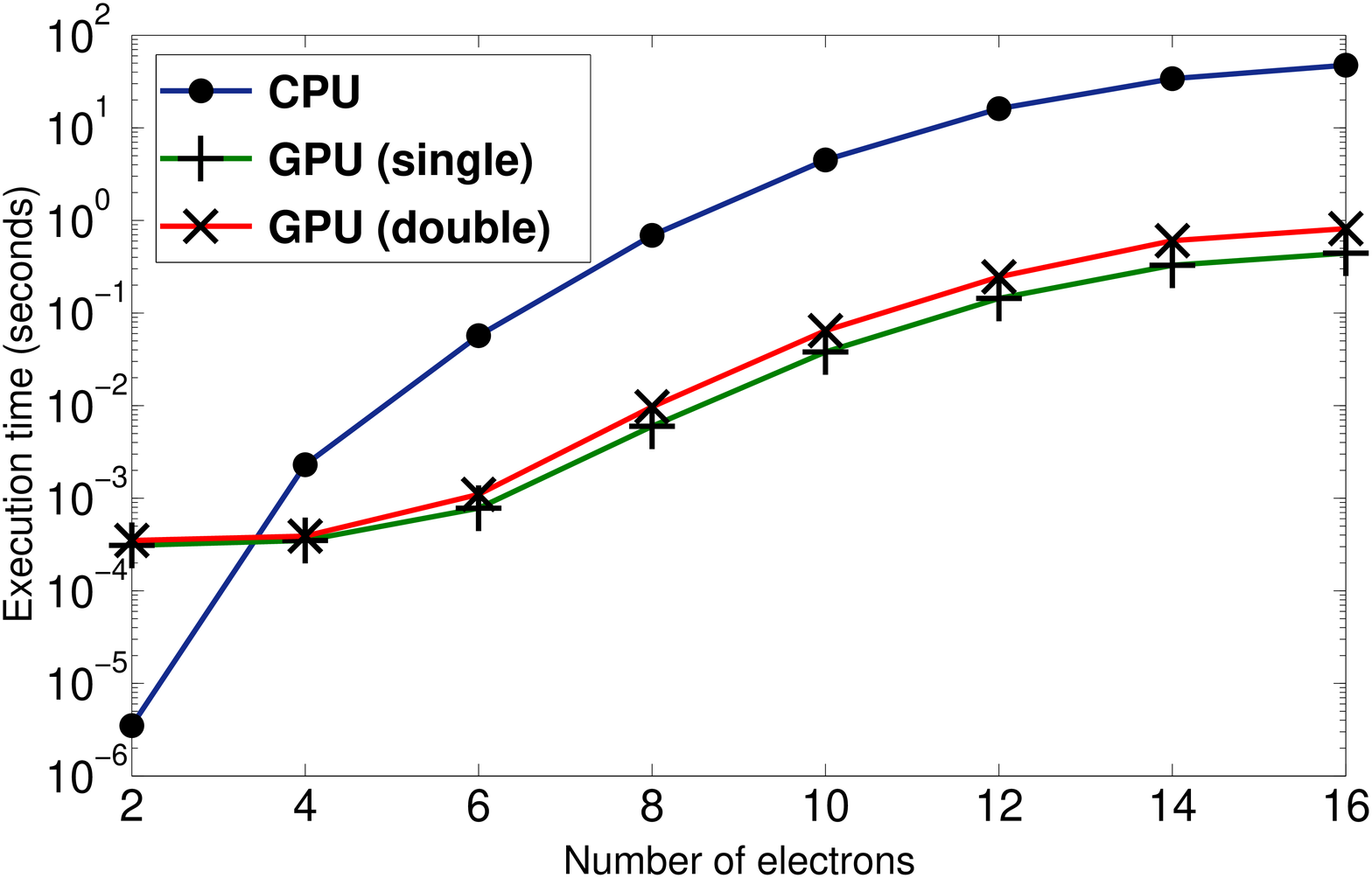} 
   \caption{The execution times of a single iteration of the Lanczos algorithm for the 2D system as a function of the number of electrons.}
   \label{fig:cpuvsgpu2D}
\end{figure}

\begin{figure}[t] 
 \includegraphics[width=0.48\textwidth]{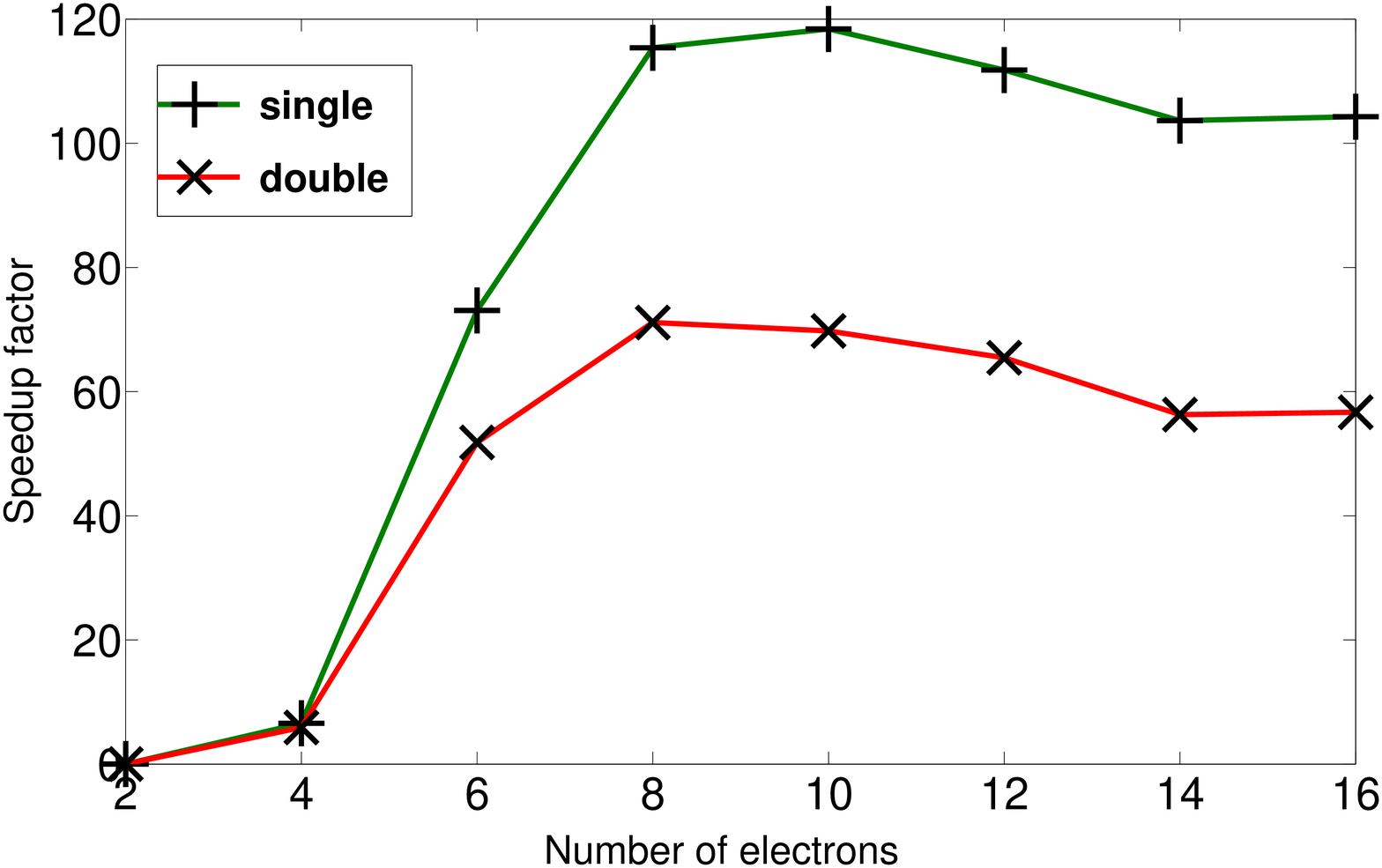} 
   \caption{The speedup factor of the GPU implementation running in single and double precision as a function of the number of electrons in the $4\times4$ square lattice. }
   \label{fig:speedup2D}
\end{figure}

In the square lattice, the situation is qualitatively similar to the 1D case. As we increase the number of electrons, the CPU execution time increases rapidly with the dimension of the Hamiltonian (Figure \ref{fig:cpuvsgpu2D}). The GPU, on the other hand, scales much better for small numbers of electrons, and starts to follow the CPU scaling as the basis grows large enough. The GPU is faster when there are more than two electrons. The maximum speedups are achieved with 10 and 8 electrons, and they are 118 and 71 for single and double precision, respectively (Figure \ref{fig:speedup2D}).

\section{Conclusions}

We have performed the exact diagonalization of the Hubbard model by implementing the Lanczos algorithm on a GPU, which was programmed with the CUDA programming model. The core of the program is the kernel that computes the result of the Hamiltonian matrix operating on a state vector, which is the most computationally demanding operation in many calculations, including the Lanczos algorithm and, for example, time propagation.

The program was benchmarked against a single-core CPU program in two lattices, a one-dimensional ring and a two-dimensional square lattice. The GPU was found to be faster when the basis size was large enough to allow for fully utilizing the resources of the GPU. In the ring, speedups of over 100 and 60 were found in single and double precision runs, respectively. In the square lattice, the corresponding speedups were over 110 and 70. The main conclusion from this work is that GPUs are well-suited for exact diagonalization calculations and significant speedups can be obtained very cost-effectively.





\bibliographystyle{model1-num-names}
\bibliography{Hubbardarticle}







\end{document}